\documentclass[letterpaper]{article} 
\usepackage{aaai2026}  
\usepackage{times}  
\usepackage{helvet}  
\usepackage{courier}  
\usepackage[hyphens]{url}  
\usepackage{graphicx} 
\usepackage{natbib}  
\usepackage{caption} 
\usepackage{algorithm}
\usepackage{algorithmic}
\usepackage{longtable}
 \usepackage[table,xcdraw]{xcolor}
\usepackage{newfloat}
\usepackage{listings}
\DeclareCaptionStyle{ruled}{labelfont=normalfont,labelsep=colon,strut=off} 
\floatstyle{ruled}
\newfloat{listing}{tb}{lst}{}
\floatname{listing}{Listing}
\usepackage{amsmath}
\usepackage{microtype}
\usepackage{inconsolata}

\title{Culture Affordance Atlas: Reconciling Object\\ Diversity Through Functional Mapping}
\author{
    Joan Nwatu \textsuperscript{\rm 1}, 
    Longju Bai \textsuperscript{\rm 1}, 
    Oana Ignat \textsuperscript{\rm 2}, 
    Rada Mihalcea \textsuperscript{\rm 1}
}
\affiliations{
    \textsuperscript{\rm 1}University of Michigan-Ann Arbor\\
    \textsuperscript{\rm 2}Santa Clara University\\

    \{jnwatu, longju, mihalcea\} @umich.edu,   oignat@scu.edu

}

\usepackage{bibentry}

\begin{document}

\maketitle

\begin{abstract}
Culture shapes the objects people use and for what purposes, yet mainstream Vision-Language (VL) datasets frequently exhibit cultural biases, disproportionately favoring higher-income, Western contexts. This imbalance reduces model generalizability and perpetuates performance disparities, especially impacting lower-income and non-Western communities. To address these disparities, we propose a novel function-centric framework that categorizes objects by the functions they fulfill, across diverse cultural and economic contexts. We implement this framework by creating the Culture Affordance Atlas, a re-annotated and culturally grounded restructuring of the Dollar Street dataset spanning 46 functions and 288 objects publicly available at \url{https://lit.eecs.umich.edu/Culture-Affordance-Atlas/index.html}. Through extensive empirical analyses using the CLIP model, we demonstrate that function-centric labels substantially reduce socioeconomic performance gaps between high- and low-income groups by a median of 6 pp (statistically significant), improving model effectiveness for lower-income contexts. Furthermore, our analyses reveals numerous culturally essential objects that are frequently overlooked in prominent VL datasets. Our contributions offer a scalable pathway toward building inclusive VL datasets and equitable AI systems.
\end{abstract}


\section{Introduction}

Culture, as reflected in data, represents people's way of life \cite{rosling2018factfulness, griswold2012cultures, tylor1871primitive}. The geographic and socioeconomic dimensions of culture profoundly influence the objects people use daily -- objects that vision-language (VL) models should accurately recognize and interpret. However, popular VL datasets suffer from persistent representational imbalances across these dimensions, despite increasing awareness of these issues \cite{longpre2024bridging, nwatu2023bridging, paullada2021data}.
Most image datasets \cite{kuznetsova2020open, deng2009imagenet, schuhmann2022laion} are object-centric and skewed toward objects common in higher-income, Western contexts. As a result, culturally distinct objects or object functions from lower-income or non-Western regions are often missing or misclassified, leading to poor model generalization.

\begin{figure}[ht]
    \centering
    \includegraphics[width=1\columnwidth]{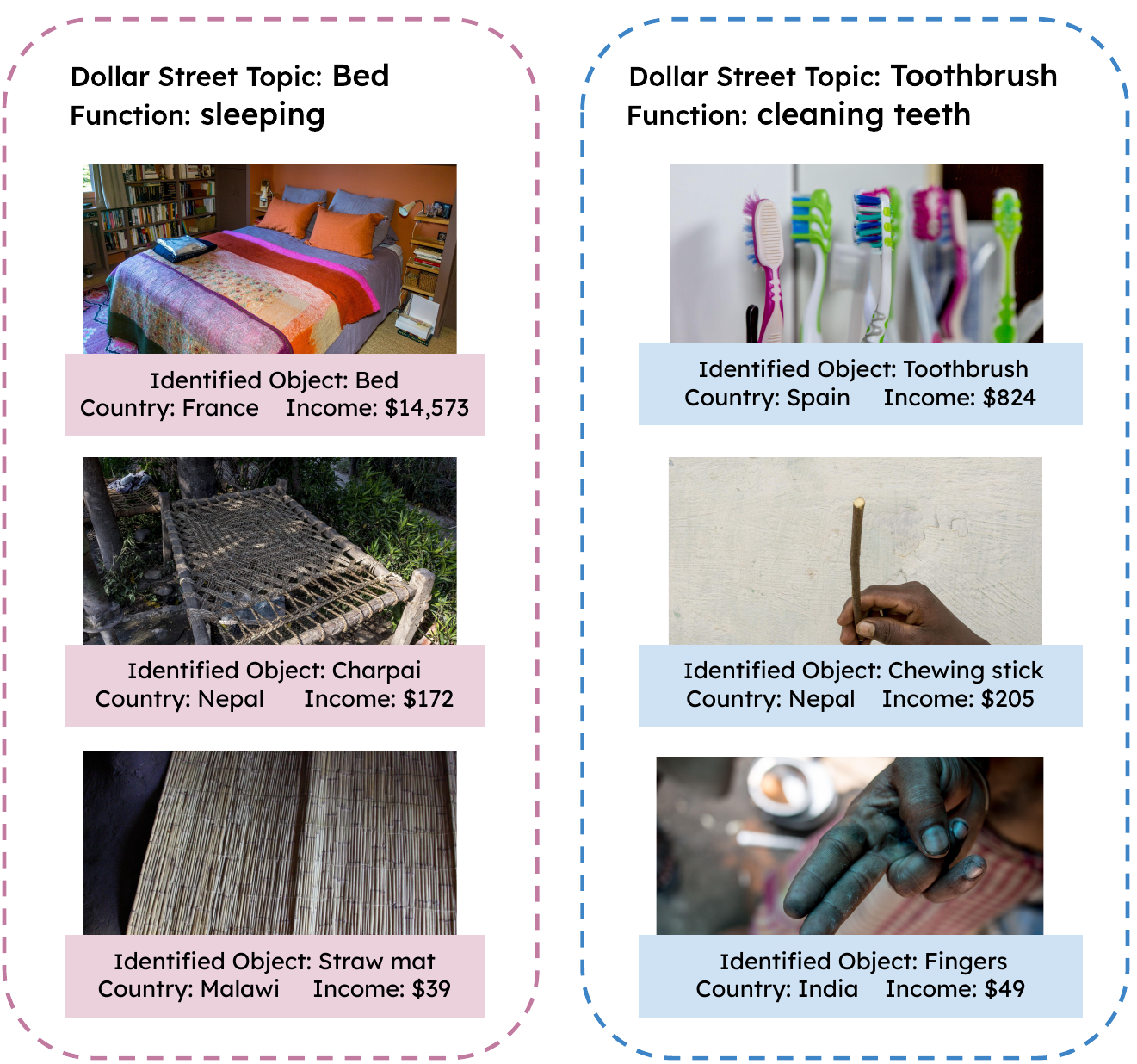}
    \caption{Various objects are identified to perform the same function across different cultures and income groups. \textit{Best viewed in color.}}
    
    \label{fig:Main}
\end{figure}

Efforts to broaden coverage such as GeoDE \cite{ramaswamy2023geode}, Dollar Street \cite{gaviria2022dollar}, and Segment Anything \cite{kirillov2023segment}, have so far made important strides, but substantial gaps remain. Large-scale crawls still filter out underrepresented data \cite{fang2023data, nwatu-etal-2025-uplifting}, and existing datasets often collapse culturally distinct artifacts under a single Western label (e.g., grouping `clay pot', `cooler box', and `refrigerator' under the label “refrigerator” in Dollar Street).

We argue that object-centric methods alone are insufficient to address these biases. We, instead, propose a function-centric lens. Across all human societies, a core set of universal activities: sleeping, cooking, cleaning, and storing are fulfilled by culturally specific objects \cite{brown2004human}. As illustrated in Figure \ref{fig:Main}, the function of `sleeping' may be served by a Western bed, a Nepali charpai, or a Malawian straw mat; `cleaning teeth' may involve a plastic toothbrush, a chewing stick, or even fingers. By re-annotating existing image collections based on such cultural affordances, we align diverse artifacts under a shared semantic framework and reveal long-tail objects overlooked in current VL taxonomies.

This paper makes the following contributions.

\begin{itemize} 

    \item We introduce a novel \textbf{function-centric framework} for constructing culturally diverse VL datasets, organizing objects by their universal human functions alongside their contextual usage. 
    \item We apply this framework to create the \textbf{Culture Affordance Atlas}, a public knowledge base derived by re-annotating the Dollar Street dataset into culturally grounded function categories, with 367 object-function pairs, each backed by at least one ethnographic citation. 
    \item We empirically demonstrate that \textbf{function-centric labeling significantly reduces disparities in VL model performance} across socioeconomic groups, highlighting practical benefits for low-income data representation. \item Our function-centric approach \textbf{uncovers numerous culturally relevant objects} frequently absent from major VL datasets underscoring its potential to bridge representational gaps.
    
\end{itemize}

Across these contributions, we demonstrate a scalable, culturally-aware path toward more inclusive VL systems. By asking \textit{` forwhat is this object used for?} rather than \textit{`what is this object?'}, we leverage a new perspective to annotate object diversity and highlight poorly represented objects in VL research.


\section{Related Work}

\subsection{Representation Bias in Vision-Language Datasets}

Multiple studies \cite{paullada2021data, shankar2017no, longpre2024bridging, ignat2024annotations, liu2024cultural} show that vision-language datasets such as LAION-5B \cite{schuhmann2022laion}, YFCC100M \cite{thomee2016yfcc100m}, COCO \cite{lin2014microsoft}, and ImageNet \cite{deng2009imagenet} contain disproportionately large amounts of data from North America and Europe, with far less representation from Africa, South Asia, and parts of Latin America. Although large datasets improve model performance, their construction often depends on institutions with substantial resources. To reduce costs, developers frequently rely on web scraping, which mirrors the distribution and priorities of the internet and tends to elevate Western high-income content \cite{birhane2021multimodal, paullada2021data, crawford2021excavating}.

Most VL datasets also use object-centric annotations based on Western taxonomies derived from ImageNet \cite{mihalcea2025ai}. This structure supports label consistency but overlooks cultural context, functionality, and alternative uses of objects \cite{nwatu-etal-2025-uplifting, nwatu2023bridging}. Recent datasets like WIT \cite{srinivasan2021wit} attempt to enrich grounding through natural language captions, yet caption-based collections still inherit platform-specific biases.

Our approach addresses these limitations by introducing structured, function-based groupings that link objects across cultures and improve representation of underrepresented household artifacts.

\subsection{Affordances, Functionality, and the Cultural Framing of Objects}

Objects are often classified by perceptual properties such as color, size, shape, texture, and material, features commonly used in deep learning for detection and recognition \cite{redmon2018yolov3, russakovsky2015imagenet}. Humans, however, frequently understand objects through their \textit{affordances}, the actions an object enables relative to a person's capabilities \cite{gibson2014theory}. This perspective highlights meaningful use rather than appearance. Anthropological work, including Brown’s Human Universals, identifies recurring domains of activity such as cooking, cleaning, sheltering, hygiene, and travel \cite{brown2004human}. These domains guide our definition of functional categories that can span visual and cultural variation.

Affordance perception is also shaped by social and cultural context \cite{costanza2020design, maier2009affordance, ye2009perceiving}. A metal basin, for instance, may afford bathing in one setting and food storage in another. In Human-Computer Interaction, overlooking such cultural variation often leads to designs that fail to engage users \cite{norman1999affordance, fayard2014affordances}.

Although affordance-based models in vision and robotics have improved generalization, especially for object manipulation \cite{hassanin2021visual, mur2023multi, castellini2011using, do2018affordancenet, bohg2013data}, they rarely consider cultural differences in object use. Our work extends affordance theory by centering cultural and functional dimensions rather than solely sensory attributes. We propose a function-based categorization framework that connects visually diverse objects performing similar daily functions across socioeconomic and geographic contexts. This approach strengthens cultural representation and reveals objects often overlooked in existing datasets.

\subsection{Improving Cultural Representation in Vision Language Models}

Efforts to improve representation in vision-language models have centered on expanding dataset diversity, adapting training methods, and refining prompts to mitigate imbalances across language, region, and socioeconomic background.

A major approach involves building datasets that better reflect underrepresented contexts. Notable examples include Segment Anything \cite{kirillov2023segment}, which offers large-scale object masks for generalization, GeoDE \cite{ramaswamy2023geode}, which evaluates model performance across geographic regions, and Dollar Street \cite{gaviria2022dollar}, which documents household items from a wide range of income levels. Although these datasets broaden coverage, they remain limited by structural issues. Many use data collection or annotation schemes shaped by Western-centric taxonomies, contain mislabeling, or lack sufficient metadata to reveal cultural distinctions. As a result, culturally important objects are often collapsed into generic categories.

Complementary work investigates prompt engineering and language adaptation to improve performance in low-resource contexts. Prior studies show that adding contextual information to prompts can increase retrieval accuracy for underrepresented data \cite{nwatu-etal-2025-uplifting, buettner2024incorporating, nguyen2023improving}. Building on this direction, we re-annotate Dollar Street using a function-centric framework. We categorize images based on the function the depicted object fulfills across cultures, assign accurate names to culturally specific items, correct prior mislabeling, and surface marginalized artifacts. These function-based groupings are then integrated into dataset labels and prompt design to address poor performance on low-income and non-Western images.

\section{Methodology}

To capture the rich diversity of artifacts serving similar functions across cultures, we propose a function-to-object labeling framework. We start with an existing dataset of cultural objects (DollarStreet) and leverage a state-of-the-art VL model (CLIP) to construct the \textbf{Culture Affordance Atlas}, a publicly accessible knowledge base documenting function-to-object mappings across diverse cultural contexts. 

\subsection{Functions}
We define \textit{functions} as the affordances of an object, that is, the specific actions or activities it enables. We explore the culturally and socially meaningful roles that objects serve across different communities, transcending their mere physical appearances. Drawing inspiration from Gibson’s theory of affordances~\cite{gibson2014theory}, and Norman’s research in human-computer interaction~\cite{norman1999affordance}, we argue that incorporating functional descriptions into object labels can improve cultural representation in AI datasets and models.

\subsection{State-of-the-art Vision-Language Model}
We use CLIP~\cite{Radford2021LearningTV} to conduct our experiments to evaluate the effectiveness of the new labels and captions on lower-income images. CLIP is a VL model trained on a large corpus of image–text pairs, enabling it to jointly embed visual and textual inputs in a shared semantic space. 
CLIP’s open availability, popularity, and relevance to zero-shot image-text retrieval and other downstream tasks, as demonstrated \citet{hessel2021clipscore} make it a suitable tool for assessing the quality and generalizability of image annotations across diverse socioeconomic contexts.


\subsection{Dollar Street Dataset}
We use the Dollar Street dataset for our re-annotation and experiments because it is geographically diverse and links household objects to reported income levels. The dataset contains 38,479 images from 63 countries across Africa, the Americas, Asia, and Europe, covering everyday objects and activities such as toothbrushes, toilet paper, and cooking. These are organized into 291 topics. Following \cite{Devries2019DoesOR}, we remove 21 subjective topics, yielding 270 objective topics for analysis.

Dollar Street also spans a wide socioeconomic range, with household incomes from $26.9 to $19,671.0 per month, adjusted for purchasing power parity. Following \cite{gaviria2022dollar}, we group incomes into geometric ranges and quartiles to support balanced comparisons. Image contributions vary across countries, from 45 images in Canada to 4,704 in India, with a median of 407 per country.



\subsection{Objects to Functions Re-annotation}
\begin{figure*}[ht]
    \centering
    \includegraphics[width=\textwidth]{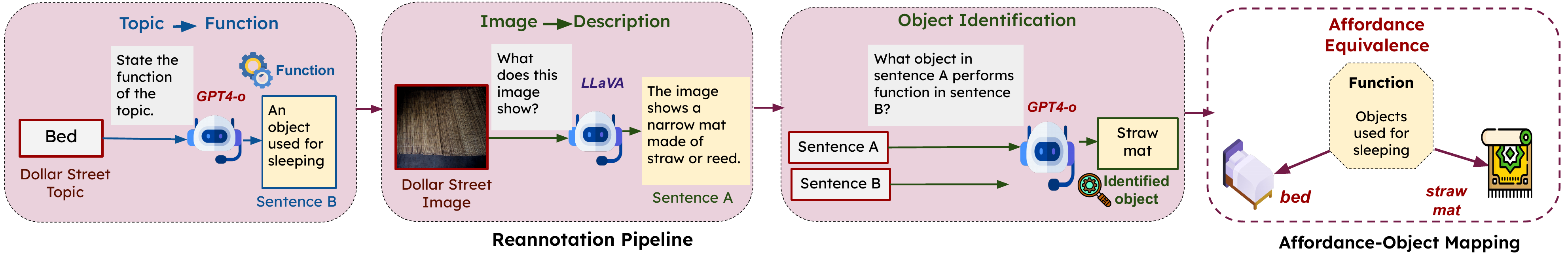}
    \caption{Re-annotation pipeline for generating functional description and identified object from original Dollar Street Image and Topic. Our Pipeline surfaces cross‐cultural affordance equivalences by mapping both image‐derived objects and topic labels to a shared functional space. \textit{Best viewed in color.}} 
    
    \label{fig:ReannotionPipe}
\end{figure*}

The Dollar Street dataset includes images as well as a topic label and an income value per each image.
We re-annotate the Dollar Street dataset to include \textit{functional descriptions} of the topic labels (e.g., `bed': `object used for sleeping') and an \textit{identified object label} which is the standard name of the object found in a given image (e.g., an image of a straw mat with the topic label `bed', will now include a functional description `object used for sleeping' and an identified object label `straw mat') as depicted in Figure \ref{fig:ReannotionPipe}.

\paragraph{Functional Descriptions.}
We generate functional descriptions for all 270 unique topic labels in Dollar Street using GPT-4o. We use the following prompt; \textit{“A bed is an object that provides a place to sleep. Following the above example, In one short sentence, state what function the object '{topic label}' does."}. We then remove the topic label mentioned in the description to get \textit{`object that provides a place to sleep'}.

\paragraph{Identified Objects.}
To identify which object present in the image is being referred to by the topic label, we tested various pipelines and continued with the pipeline that produced better outcomes.
We use LLaVa to caption the images \textit{(prompt- ``USER: [image] What does this image show? ASSISTANT:")} and then use GPT-4 to extract the name(s) of the objects present in the image caption that fulfills the associated functional description for the image with the prompt: \textit{'You are given two sentences. One: {function} Two: {description} Mention the name of the object or objects referred to in both sentences'} where sentence one is the function and sentence two is the image caption.

\paragraph{Quality Control.}
We randomly check for incorrect entries to rewrite and manually reannotated 1458 identified object names including those that were previously listed as \textit{NaN}, \textit{None} or \textit{Not identifiable}. 

For the functional descriptions, we run a quality assessment check with 21 participants across 7 countries (China, India, Ethiopia, Nigeria, United States, Romania, and Russia) selected due to their representation of the 4 continents present in Dollar Street and availability of annotators. For each of these countries, 30 questions are formulated using randomly selected images\footnote{All image selections for the user study were drawn by shuffling the dataset using pandas.DataFrame.sample with a fixed random\_state (set to 42) to ensure reproducibility.} and their respective functional descriptions. 
For each country, three native participants evaluate whether the provided functional description accurately describes the image, responding with a binary (yes/no) judgment.

Using responses from the assessment, we calculate the percentage validation for each country and the overall average. We report an overall score of 90\% and show in that all scores across the seven countries are above 85\% (Table \ref{tab:Qualitycheck} . This indicates a high correspondence between the Dollar Street images and the generated functional descriptions. We also include computed multiple inter-annotator agreement metrics in the appendix Table \ref{tab:interannotator}.

\begin{table}[ht]
\centering
\resizebox{0.5\columnwidth}{!}{%
\begin{tabular}{l|r}
\textbf{Country} & \multicolumn{1}{l}{\textbf{\% validated}} \\ \hline
China & 87.78 \\
India & 90.00 \\
Ethiopia & 91.11 \\
Nigeria & 86.67 \\
United States & 87.78 \\
Romania & 96.67 \\
Russia & 90.00 \\ \hline
\textbf{Average} & \textbf{90.00} \\
\end{tabular}%
}
\caption{Quality control for generated functional descriptions}
\label{tab:Qualitycheck}
\end{table}


\section{Culture Affordance Atlas.}
Among the various functions humans perform (e.g., writing, cooking, creating art), we prioritize those universally documented by \citet{brown2004human} in \textit{Human Universals} which is an anthropological theory that examines shared human characteristics and behaviors across cultures. We initially select seven human universal categories: hygienic care, aesthetics, cooking, diurnality, environmental adjustments, healing the sick, and visiting (details in Appendix Table \ref{tab:HumanUniversals}).
We map functional descriptions and associated object names identified from the Dollar Street re-annotation pipeline (Figure \ref{fig:ReannotionPipe}) to functions within these overarching categories. For each primary category (e.g., cooking), we assign relevant functions and document the corresponding objects identified within the Dollar Street dataset.
To ensure rigorous grounding, each entry is supported by at least one published, verifiable source. Approximately 98\% of references originate from ethnographic publications accessed through the eHRAF World Cultures database\footnote{https://ehrafworldcultures.yale.edu/}, and Google scholar for the remaining 2\%. The initial edition of the Culture Affordance Atlas  comprises 367 entries, representative images, and metadata from Dollar Street, all publicly accessible at \url{https://lit.eecs.umich.edu/Culture-Affordance-Atlas/index.html}.
\paragraph{General Statistics.}
Culture Affordance Atlas currently includes 367 entries spanning 7 categories, 46 functions, and 288 unique objects. Each function-object pair (e.g., teeth cleaning-toothbrush) includes a frequency count derived from contextual image annotations (Table \ref{tab:CAAstats}). Summary statistics and representative examples across the categories of the Atlas are provided in Table \ref{tab:CAAstats}.

\begin{table}[ht]
\centering
\resizebox{0.95\columnwidth}{!}{%
\begin{tabular}{l|r|r|r|l}
\textbf{\begin{tabular}[c]{@{}l@{}}Culture Affordance \\ Category\end{tabular}} & \multicolumn{1}{l|}{\textbf{\#Func}} & \multicolumn{1}{l|}{\textbf{\#Obj}} & \multicolumn{1}{l|}{\textbf{\begin{tabular}[c]{@{}l@{}}Function \\ example\end{tabular}}} & \textbf{\begin{tabular}[c]{@{}l@{}}Object \\ example\end{tabular}} \\ \hline
\begin{tabular}[c]{@{}l@{}}Practicing \\ hygiene\end{tabular} & 19 & 96 & \begin{tabular}[c]{@{}r@{}}cleaning \\ teeth\end{tabular} & toothbrush \\
Beautifying & 3 & 40 & \begin{tabular}[c]{@{}r@{}}adorning \\ the body\end{tabular} & necklace \\
Cooking & 7 & 60 & \begin{tabular}[c]{@{}r@{}}providing heat\\  for cooking\end{tabular} & \begin{tabular}[c]{@{}l@{}}charcoal \\ stove\end{tabular} \\
\begin{tabular}[c]{@{}l@{}}Maintaining a \\ diurnal cycle\end{tabular} & 3 & 16 & sleeping & mat \\
\begin{tabular}[c]{@{}l@{}}Adjusting to\\  the environment\end{tabular} & 11 & 85 & providing light & oil lamp \\
Healing & 2 & 16 & \begin{tabular}[c]{@{}r@{}}providing \\ medication\end{tabular} & inhaler \\
Traveling & 1 & 10 & transporting & bicycle
\end{tabular}%
}
\caption{Statistics for the Culture Affordance Atlas.}
\label{tab:CAAstats}
\end{table}

\paragraph{Label Misalignment and Object Diversity within Functions.}
Through our re-annotation pipeline, we identify discrepancies between original Dollar Street topics and actual object use. A notable percentage of images display mismatches between original topic labels and identified objects, averaging \textbf{38.25\%} misalignment across topics.\footnote{We prompt GPT4o to identify mismatches by comparing Dollar Street Topics with the identified objects from our pipeline.} Functions such as ``bathing'' and ``clothes washing'' demonstrate significant object diversity, characterized by visually disparate artifacts.

\begin{figure}[ht]
    \centering
    \includegraphics[width=0.95\columnwidth]{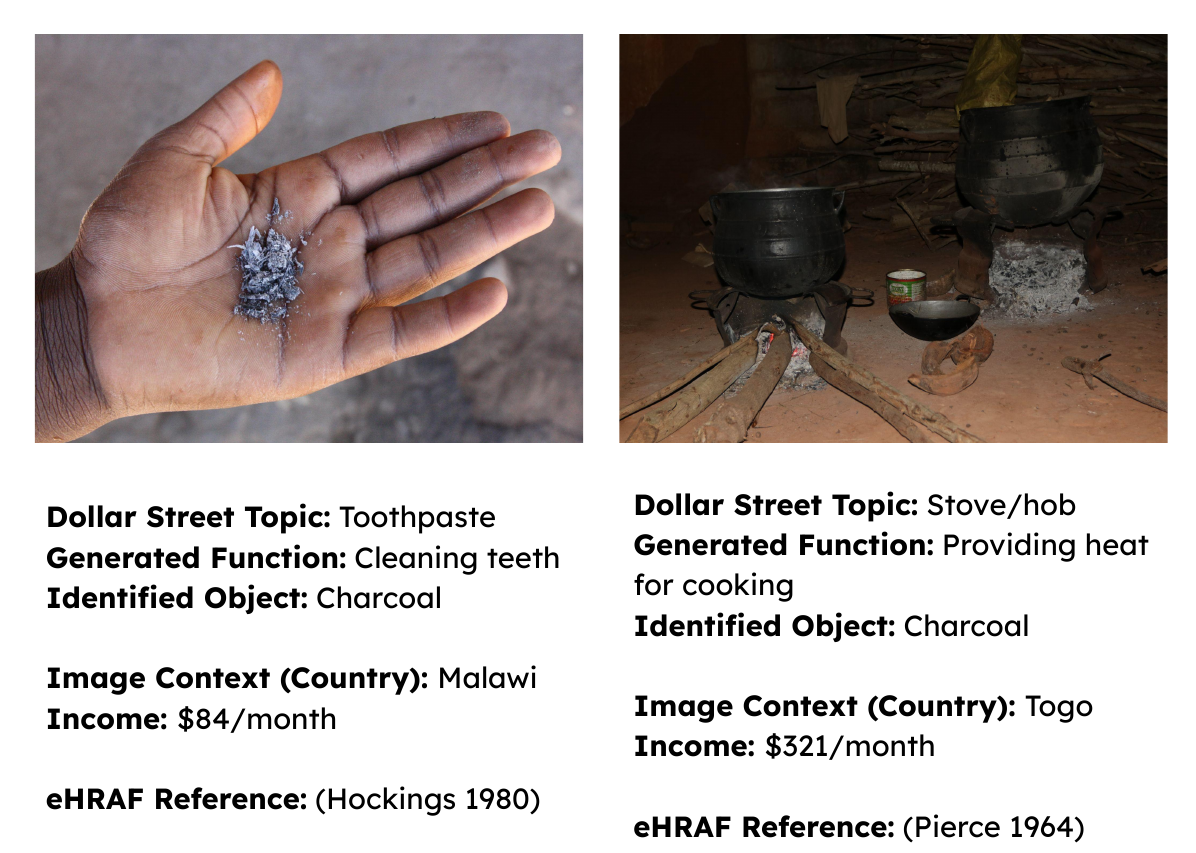}
    \caption{Charcoal use across functions. \textit{Best viewed in color.}} 
    \label{fig:coal}
\end{figure}

\paragraph{Object Use Across Functions.}
Many objects fulfill multiple functions. For example, ``bowls'' appear in contexts including ``clothes washing'', ``hand washing'', ``food serving'', and ``food preparation''. Object-function pairs demonstrating alternative uses, differing from an object's primary function, highlight the versatility and improvisational use of everyday items. ``Charcoal'' exemplifies this phenomenon: primarily used as ``fuel'', it also serves ``dental hygiene'' functions (Figure \ref{fig:coal}). Such objects, termed \textit{contextually niche}, have dominant uses alongside lesser-known but culturally significant secondary functions. The Culture Affordance Atlas enhances visibility for these secondary functions by documenting relevant contexts and ethnological references.

\paragraph{Long-Tail Objects.}

Vision-language research often focuses on common object categories, leaving niche or culturally specific items underrepresented and contributing to a long-tail distribution. Accounting for these overlooked objects is important for building more inclusive and robust models.

We identify long-tail objects in the Culture Affordance Atlas that appear infrequently in our re-annotation. Using the 100 least frequent items (e.g., calabash, earthen pot), we examine their presence across seven widely used VL datasets selected for their broad object coverage and prominence: ImageNet \cite{deng2009imagenet}, COCO \cite{lin2014microsoft}, Open Images \cite{kuznetsova2020open}, LVIS \cite{gupta2019lvis}, GeoDE \cite{ramaswamy2023geode}, YFCC100M Entity (Artifacts) \cite{li2017learning}, and Dollar Street \cite{gaviria2022dollar}. To ensure consistent comparison, we normalize labels using lowercasing, lemmatization, and fuzzy matching with a threshold of 0.8.


\begin{table}[ht]
\centering
\resizebox{0.95\columnwidth}{!}{%
\begin{tabular}{l|r|r|r|r}
\multicolumn{1}{c|}{\textbf{Dataset}} & \multicolumn{1}{c|}{\textbf{\begin{tabular}[c]{@{}c@{}}Total\\ classes\end{tabular}}} & \multicolumn{1}{c|}{\textbf{\begin{tabular}[c]{@{}c@{}}Long‐tail\\ covered\end{tabular}}} & \multicolumn{1}{c|}{\textbf{Missing}} & \multicolumn{1}{c}{\textbf{\begin{tabular}[c]{@{}c@{}}Coverage\\ (\%)\end{tabular}}} \\ \hline
ImageNet & 1000 & 8 & 92 & 8\% \\
COCO & 80 & 2 & 98 & 2\% \\
OpenImages & 19868 & 61 & 39 & 61\% \\
LVIS & 1230 & 28 & 72 & 28\% \\
GeoDE & 40 & 0 & 100 & 0\% \\
\begin{tabular}[c]{@{}l@{}}YFCC100M\\ Entity-Artifacts\end{tabular} & 325 & 1 & 99 & 1\% \\
Dollar Street & 291 & 5 & 95 & 5\%
\end{tabular}%
}
\caption{Long-tail Objects.  Percentage of our 100 long‐tail objects are present in each popular vision-language dataset’s official class list.}
\label{tab:universal_longtail}
\end{table}

Our results in Table \ref{tab:universal_longtail} reveal that only OpenImages, due to its extensive label set ($\approx$20k classes), covers more than half of these rare objects. We further identify 34 universally absent objects, with 33 appearing in only one dataset. This underscores persistent gaps in dataset representation, highlighting the need for comprehensive documentation and integration of culturally diverse and niche artifacts within VL research.
A breakdown of these 100 long-tail objects and their presence across VL datasets is available in the appendix Table \ref{tab:longtail_details2}.

\section{Experiments}
\subsection{The Effects of Function Captions on VL models.}

\begin{figure}[ht]
    \centering
    \includegraphics[width=1\columnwidth]{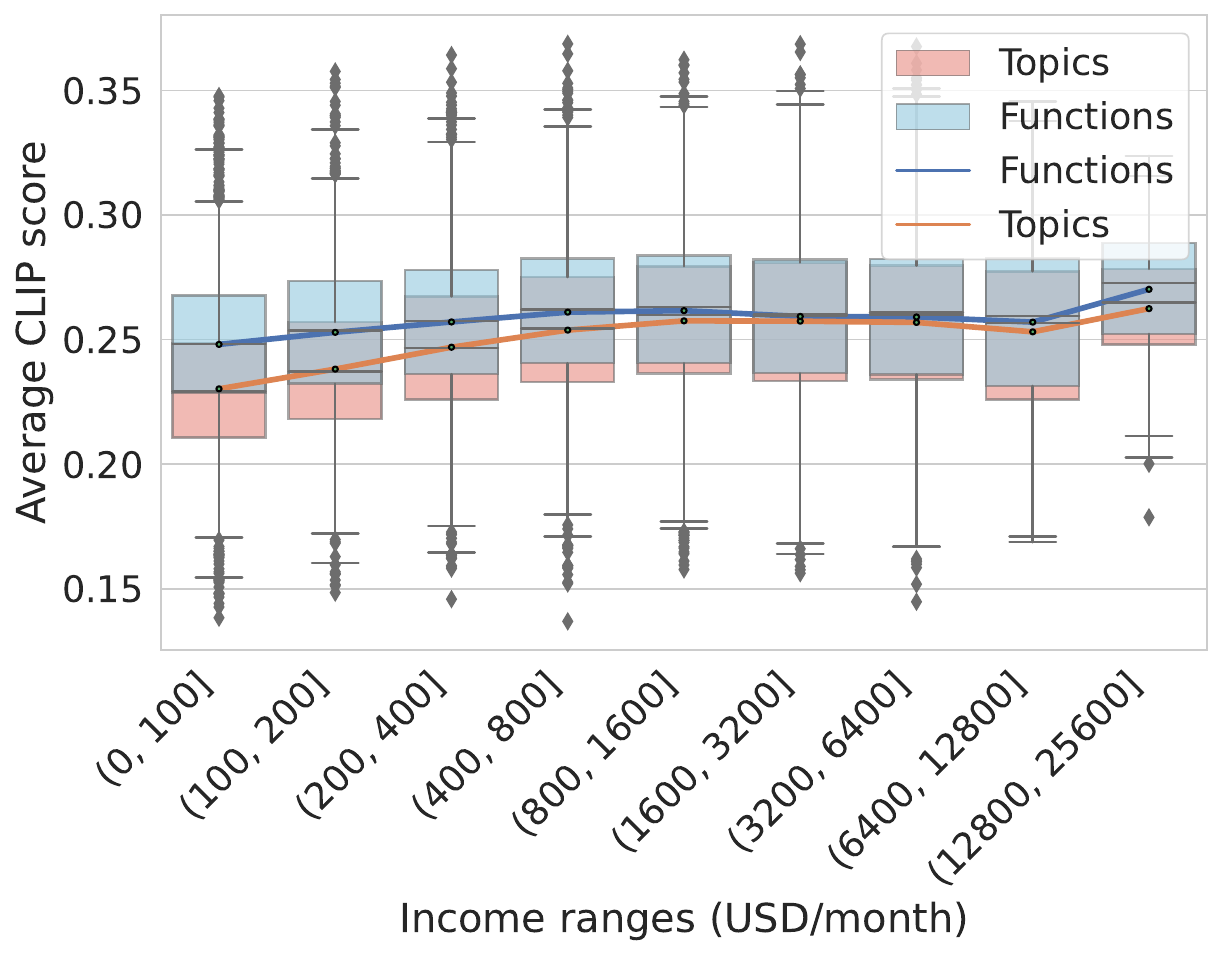}
    \caption{Comparison of CLIP alignment scores between Topic-Image (red) and Function-Image (blue) pairs across Dollar Street images from varying income levels. Trend lines indicate mean scores per income bin. The slope of each line reflects the extent of the digital divide. \textit{Best viewed in color.}}
    
    \label{fig:TopicvsFunctionBox}
\end{figure}

We investigate the impact of function-based captions on VL model performance through two distinct experiments comparing original Dollar Street topic labels against corresponding function descriptions generated via our re-annotation pipeline (Figure \ref{fig:ReannotionPipe}).
In the first experiment, termed the \textit{CLIP association test}, we compute cosine similarities between Dollar Street image embeddings and text embeddings of both topics (baseline) and functions. Images are grouped according to geometric income bins as described in \cite{gaviria2022dollar}.
In the second experiment, we evaluate image retrieval performance. Specifically, we calculate CLIP scores between all images and each Dollar Street topic to derive topic-image associations, repeating the process using corresponding function captions to obtain function-image associations. We then measure retrieval performance using Recall: we retrieve the top \textit{N} images based on CLIP scores—where \textit{N} equals the number of ground-truth images—and compare these retrieved images against the ground-truth (i.e., number of true predictions / \textit{N}).
Below, we present key findings from these experiments.

\paragraph{Effects of Functions on Income Performance Gaps.}

Figure \ref{fig:TopicvsFunctionBox} compares CLIP alignment scores for images grouped by income, using functions (blue) and topics (red). Trend lines depict the average CLIP scores per income bin. Consistent with findings from \cite{nwatu2023bridging}, CLIP scores generally increase with income, indicating a performance gap favoring higher-income contexts. However, the function-based scores produce a flatter trajectory compared to topics. Linear regression confirms this visual observation, with slopes of 0.002 for functions versus 0.004 for topics, indicating that function-based captions significantly reduce performance disparities.

For each of the 270 topics, we computed the difference in recall between high (rich and up-mid) and low-income (poor and low-mid) image sets under topic-only and function-based prompts, then took the per-topic gap reduction (\(\Delta gap\)) \footnote{Per topic, gap in Topic prompt - gap in Function Prompt}. A Wilcoxon signed-rank test \cite{woolson2005wilcoxon} shows that (\(\Delta gap\)) is significantly greater than zero (median = 0.06, p = 1.62e-17), confirming that function-based prompting reliably narrows socioeconomic performance gaps.

\begin{figure}[ht]
    \centering
    \includegraphics[width=1\columnwidth]{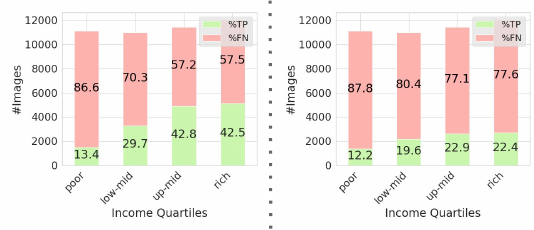}
    \caption{CLIP Recall across all images using Topic-Image (left) and Function-Image (right) alignment scores. We report the percentage of true positives (``recognized'' images) and false negatives (``forgotten'' images) for each income quartile. Function-Image Recall shows less variation across income levels compared to Topic-Image Recall, indicating greater robustness to income-based distribution shifts. \textit{Best viewed in color.}}
    \label{fig:TopicvsFunctionBar}
\end{figure}

\begin{figure}[ht]
    \centering
    \includegraphics[width=1\columnwidth]{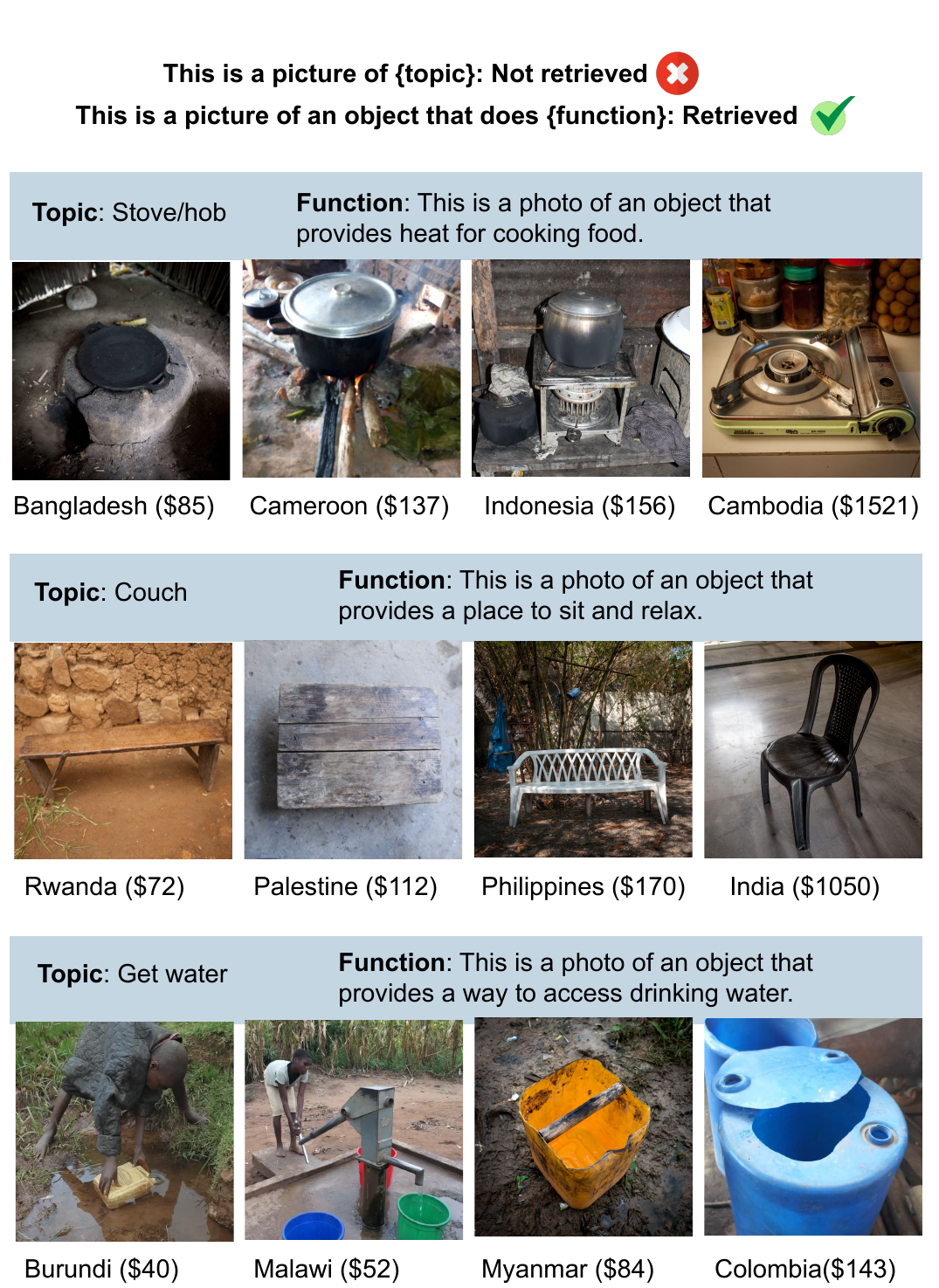}
    \caption{Qualitative analysis of Images ‘forgotten’ during retrieval with \textit{topic prompt} but successfully retrieved with the \textit{function prompt}
    \textit{Best viewed in color.}}
    
    \label{fig:Qual_function}
\end{figure}

\begin{figure}[ht]
    \centering
    \includegraphics[width=0.90\columnwidth]{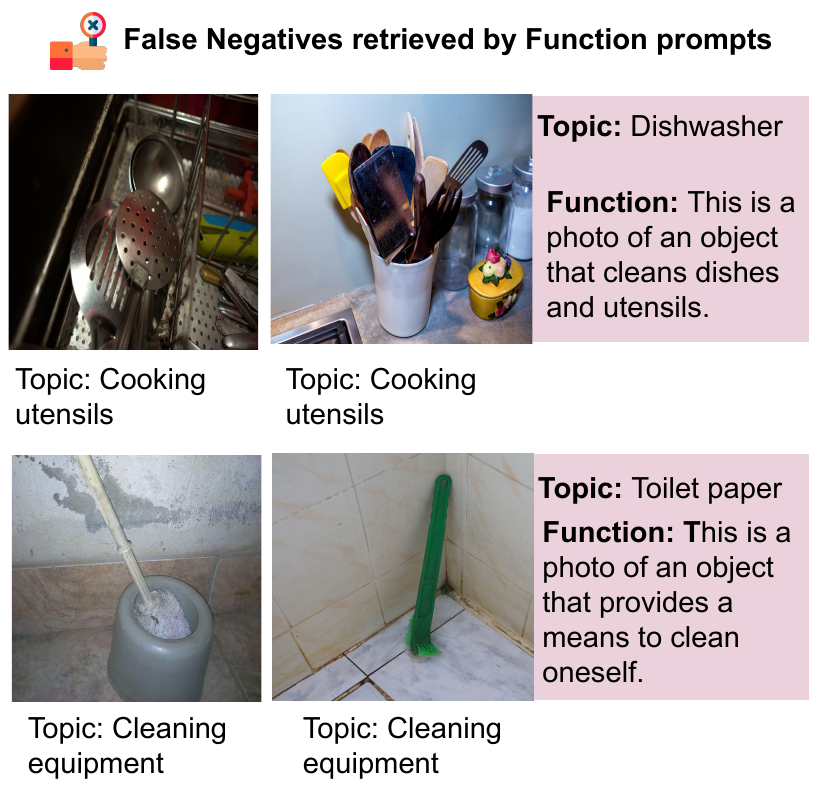}
    \caption{Qualitative analysis of false positives retrieved by \textit{function prompts.} \textit{Best viewed in color.}}
    
    \label{fig:TopicvsFunctionQuals}
\end{figure}

Figure \ref{fig:TopicvsFunctionBar} illustrates Recall results across income quartiles (poor, low-mid, upper-mid, and rich). Echoing the trends observed previously, functions yield less pronounced Recall disparities across income groups. Nevertheless, overall Recall scores for functions are lower compared to topics, suggesting trade-offs when employing functional descriptions in retrieval tasks. This phenomenon is explored further in the next subsection.

\paragraph{Tradeoffs and Challenges in Functions for VL tasks.} 

Functions prompt CLIP to broaden retrieval and include culturally diverse, non-standard objects that topic prompts often miss. As shown in Figure \ref{fig:Qual_function}, various forms of ``stoves'', ``couches'', and ``water sources'' from across cultures are correctly retrieved when framed by their function rather than their topic.

However, while function-based labels help reduce digital divides and equalize performance across socioeconomic contexts, they also decrease retrieval accuracy. We explore this trade-off qualitatively, by examining false positives produced by function prompts.
We find that misinterpretation of functions by the model can lead to the retrieval of contextually incorrect objects, for example, the tendency of CLIP to emphasize nouns independently rather than interpreting full sentence contexts \cite{castro2023scalable}. In Figure \ref{fig:TopicvsFunctionQuals}, the function prompts retrieve dishware and cultery instead of ``objects that clean dishes'', and retrieve images related to general cleaning of the toilets but miss the intended context of personal hygiene in the prompt. Effective use of functions thus requires careful crafting of prompts that are precise yet sufficiently inclusive to capture reasonable diversity (e.g., preferring \textit{object that cleans dishes} over \textit{machine that cleans dishes}).

    

\paragraph{Integrating Topics and Functions.}
Motivated by these insights, we explore integrating topics with their respective functions into combined prompts. For example, the topic \textit{dishwasher} and its function \textit{an object that cleans dishes and utensils} merge into the prompt \textit{A dishwasher that cleans dishes and utensils}.

Repeating our experiments with these combined function-topic prompts, we compare performance against standalone topics (see the appendix (Figure \ref{fig:TopicvsFunctionTopicBox}) and (Figure \ref{fig:TopicvsFunctionTopicBar}) for more details).   Similar to function prompts, function-topic prompts eliminate performance disparities across income levels, achieving a near-zero slope (-0.0002) in CLIP association tests, with particularly notable performance gains for lower-income bins. Consequently, recall scores improve markedly for lower income quartiles, although slight reductions occur in higher income quartiles.

\section{Generalizability}
We confirm our findings on the reduction of performance disparity across income across vision-language models by repeating our experiments on the siglip2-so400m-patch14-384 model \cite{tschannen2025siglip2multilingualvisionlanguage}. Consistent with our findings on CLIP, we find that function prompts significantly reduce the high–low income performance gap by 11\% ($p = 5.9e-11$, Wilcoxon signed-rank test). Plots for the siglip2-so400m-patch14-384 model alignment scores and recall can be found in the appendix Fig \ref{fig:siglipTopicvsFunctionBox} and \ref{fig:siglipTopicvsFunctionBar}.

\section{Lessons Learned}

We highlight key insights learned from our findings and present them below.

\paragraph{Cultural and socioeconomic biases persist in mainstream VL datasets.} 
Our analysis reveals that 34\% of our curated long-tail objects are entirely absent from seven widely used VL datasets, while 38.25\% of Dollar Street topics suffer from label misalignment. These findings expose a critical blind spot in dataset construction: culturally specific artifacts are routinely omitted or mislabeled. Left unaddressed, such gaps risk entrenching cultural erasure, amplifying systemic bias, and further marginalizing underrepresented communities in AI applications.

\paragraph{Function-centric labeling effectively reduces the socioeconomic performance gap.}
We find that function-centric labeling reduces the recall gap between high and low-income image sets by a median of 6 percentage points in CLIP and 11 percentage points in SigLIP2 ($p < 5.9e-11$). These results demonstrate that focusing on what objects afford, rather than how they are named, leads to more equitable model performance across income levels. Notably, designing precise, yet inclusive function prompts is crucial, as overly broad prompts reduce retrieval accuracy, while overly specific prompts fail to capture object diversity.

\paragraph{Systematic documentation is necessary for improving object representation in VL datasets.}
Following the recommendation from \citet{nwatu2023bridging} to annotate diversity and subjectivity in datasets, we present the Culture Affordance Atlas as an effort to reconcile object appearance diversity within labels and inclusively categorize them. We call for more interdisciplinary research efforts involving the AI community and domain experts toward representative data annotation.


\section{Conclusion}

In this paper, we demonstrated that culturally informed, function-based labeling substantially reduces representational disparities in vision-language models. By introducing the Culture Affordance Atlas, we systematically captured culturally diverse object-function relationships, addressing critical gaps in mainstream VL datasets. Our empirical analysis using CLIP showed that integrating object functions with traditional labeling significantly improves performance equity across socioeconomic contexts. This work demonstrates that adopting culturally aware frameworks such as function-centric annotation in dataset construction facilitates the development of inclusive AI systems capable of reliably serving diverse global communities. Our analysis code, the re-annotated Dollar Street dataset and the Culture Affordance Atlas webpage are available at \url{https://github.com/MichiganNLP/Culture-Afford-Analysis}.

\section{Limitations}

\textbf{Culture Affordance Atlas Object Documentation:} Each object–function pair in the Atlas is supported by at least one verifiable published source confirming its real-world use. Due to resource constraints, we did not exhaustively document all possible citations per culture. Most entries (89\%) include one credible reference, the first reliably identified one, while 11\% contain multiple sources. We position this Atlas as a foundational layer of cultural grounding that future work can expand through additional eHRAF searches, museum and archival materials, and community contributions. Current references should be viewed as initial guides for deeper inquiry.

\textbf{Long-Tail Objects:} The 34 long-tail objects represent items that did not match any class in ImageNet, COCO, OpenImages, LVIS, GeoDE, YFCC100M Entity (Artifacts), or Dollar Street after normalization and fuzzy matching (threshold 80). Because synonym variation may occur, we provide a `Notes' column in the supplemental table to flag potential ambiguities.

\textbf{Validation Scope:} Our validation involved 21 participants across seven countries, providing geographically diverse and cross-cultural assessment. Nonetheless, broader participation would further strengthen the robustness of the validation.

\section*{Acknowledgments}
We thank the anonymous reviewers for their constructive feedback. We are also grateful to the Language and Information
Technologies (LIT) lab members at the University
of Michigan for their insightful discussions and
feedback during the project’s early stages. This project was partially funded by an award from OpenAI. 
Any opinions, findings, and conclusions or recommendations expressed in this material are those of the authors and do not necessarily reflect the views of Open AI.

\bigskip

\bibliography{aaai2026}

\section{Reproducibility Checklist}

\begin{itemize}
    \item Does this paper include a conceptual outline and/or pseudocode description of AI methods introduced? \textbf{(Yes)}
    \item Does this paper clearly delineate statements that are opinions, hypothesis, and speculation from objective facts and results? \textbf{(Yes)}
    \item Does this paper provide well marked pedagogical references for less-familiare readers to gain background necessary to replicate the paper? \textbf{(Yes)}
    \item Does this paper make theoretical contributions? \textbf{(No)}
    \item Does this paper rely on one or more datasets? \textbf{(Yes)}
    \item Is a motivation given for why the experiments are conducted on the selected datasets?  \textbf{(Yes)}
    \item Are all novel datasets introduced in this paper are included in a data appendix? \textbf{(Yes)}
    \item Are all novel datasets introduced in this paper will be made publicly available upon publication of the paper with a license that allows free usage for research purposes? \textbf{(Yes)} 
    \item Are all datasets drawn from the existing literature (potentially including authors’ own previously published work) are publicly available? \textbf{(Yes)}
    \item Are all datasets that are not publicly available are described in detail, with explanation why publicly available alternatives are not scientifically satisfactory? \textbf{(N/A)}
    \item Does this paper include computational experiments? \textbf{(Yes)}
    \item Does this paper state the number and range of values tried per (hyper-) parameter during development of the paper, along with the criterion used for selecting the final parameter setting? \textbf{(N/A)}
    \item Does this paper ensure that any code required for pre-processing data is included in the appendix? \textbf{(Yes)}, in Methodology and code appendix \url{https://anonymous.4open.science/r/Culture-Afford-Analysis-4094/}
    \item Does this paper include all source code required for conducting and analyzing the experiments in a code appendix? \textbf{(Yes)}
    \item Will all source code required for conducting and analyzing the experiments will be made publicly available upon publication of the paper with a license that allows free usage for research purposes? \textbf{(Yes)} 
    \item Will all source code implementing new methods have comments detailing the implementation, with references to the paper where each step comes from? \textbf{(Yes)}
    \item If an algorithm depends on randomness, then the method used for setting seeds is described in a way sufficient to allow replication of results? \textbf{(Yes)}
    \item Does this paper specify the computing infrastructure used for running experiments (hardware and software), including GPU/CPU models; amount of memory; operating system; names and versions of relevant software libraries and frameworks? \textbf{(Yes)} in the anonymous code appendix.
    \item Does this paper formally describe the evaluation metrics used and explain the motivation for choosing these metrics? \textbf{(Yes)}
    \item Does this paper state the number of algorithm runs used to compute each reported result. \textbf{(Yes)} Due to the nature of our evaluation setup, our Recall and CLIP-association experiments return the same values across multiple runs.
    \item Does the analysis of experiments go beyond single-dimensional summaries of performance (e.g., average; median) to include measures of variation, confidence, or other distributional information. \textbf{(Yes)}
    \item Is the significance of any improvement or decrease in performance judged using appropriate statistical tests (e.g., Wilcoxon signed-rank)? \textbf{(Yes)}
    \item Does this paper list all final (hyper-)parameters used for each model/algorithm in the paper’s experiments. \textbf{(Yes)}
\end{itemize}

\appendix
\section{Appendix}

\subsection{Human Universals used in the Culture Affordance Atlas}

\begin{table*}[ht]
\resizebox{\textwidth}{!}{%
\begin{tabular}{l|l|l|l|l}
\textbf{Original  Human Universal (Brown)} & \textbf{Gerund Conversion} & \textbf{Culture Affordance Category} & \textbf{Notes} & \textbf{References} \\ \hline
Hygienic Care & Practicing hygiene & Practicing hygiene & The basic need for health gives rise to hygiene. & \begin{tabular}[c]{@{}l@{}}Human Universals by \\ Donald E. Brown 1991 (page 67)\end{tabular} \\
Aesthetics & Beautifying & Beautifying & \begin{tabular}[c]{@{}l@{}}Emotion producing entities and activities (we focus on \\ beautification of human body and environment)\end{tabular} & \begin{tabular}[c]{@{}l@{}}Human Universals by \\ Donald E. Brown 1991 (page 115)\end{tabular} \\
Cooking & None & Cooking & Food preparation using fire is universal and well-dated & \begin{tabular}[c]{@{}l@{}}Human Universals by \\ Donald E. Brown 1991 (page 95)\end{tabular} \\
Diurnality & Maintaining a diurnal cycle & Maintaining a diurnal cycle & Engaging in daytime activity and sleeping at night & \begin{tabular}[c]{@{}l@{}}Human Universals by \\ Donald E. Brown 1991 (page 139)\end{tabular} \\
Environment, adjustments to & Adjusting to the environment & Adjusting to the environment & Material adjustments to living environments & \begin{tabular}[c]{@{}l@{}}Human Universals by \\ Donald E. Brown 1991 (page 136)\end{tabular} \\
Healing of the sick (or attempting to) & Healing & Healing & Healing the sick, use of medicines & \begin{tabular}[c]{@{}l@{}}Human Universals by \\ Donald E. Brown 1991 (page 139)\end{tabular} \\
Visiting & Traveling & Traveling & \begin{tabular}[c]{@{}l@{}}Visiting kin or others who dwell elsewhere (we focus \\ on the  movement/travel necessary for a visit to occur)\end{tabular} & \begin{tabular}[c]{@{}l@{}}Human Universals by \\ Donald E. Brown 1991 (page 139)\end{tabular}
\end{tabular}%
}
\caption{The 7 human universals that make up the overarching function categories for the Culture Affordance Atlas}
\label{tab:HumanUniversals}
\end{table*}

\paragraph{Universal Longtail}
The list of 34 universal long-tail objects are as follows [`antiseptic',
 `balm',
 'bamboo enclosure',
 `bath drain',
 `bindi',
 `bladeless fan',
 `box of medication',
 `brick oven',
 `charcoal stove',
 `clay pot',
 `clay stove',
 `clay/brick powder',
 `earthern pot',
 `electric heater',
 `floor mat',
 `fragrance',
 `furnace',
 `jerrycan',
 `landfill',
 `laundry pod',
 `led light',
 `medicinal liquid',
 `ointment',
 `panty liner',
 `charpai',
 `shaving foam',
 `skylight',
 `stone stove',
 `tooth powder',
 `tooth soap',
 `tree trunk',
 `vent pipe',
 `ventilation hole',
 `washboard']

\paragraph{Longtail objects present in only 1 of the 7 datasets}
They include the following 33 objects; [`calabash',
 `keycard',
 `collage',
 `van',
 `hand sanitizer',
 `water tank',
 `inhaler',
 `rack',
 `gas stove',
 `scooter',
 `cistern',
 `rosary',
 `mat',
 `brush',
 `shed',
 `mural',
 `hand fan',
 `statue',
 `river',
 `rock',
 `massage chair',
 `faucet',
 `sinkhole',
 `mascara',
 `eyeliner',
 `tampon',
 `cane',
 `tent',
 `bean bag chair',
 `water well',
 `potted plant',
 `rake',
 `cabin']

\paragraph{Detailed Table with Longtail Objects from the Culture Affordance Atlas}

\begin{table*}[ht]
\centering
\resizebox{\textwidth}{!}{%
\begin{tabular}{c|c|c|c|c|c}
\hline
\textbf{\begin{tabular}[c]{@{}c@{}}Human Universal \\ Affordance\end{tabular}} & \textbf{Funtional Name} & \textbf{Objects} & \textbf{Coverage} & \textbf{Datasets Covered} & \textbf{Notes/Possible rephrase} \\ \hline
Adjusting to the environment & cooling living spaces & bladeless fan & \cellcolor[HTML]{F4CCCC}Universally missing & ['None'], &  \\
Adjusting to the environment & providing heat & electric heater & \cellcolor[HTML]{F4CCCC}Universally missing & ['None'], & heater/risks loss of granularity \\
Adjusting to the environment & providing light & skylight & \cellcolor[HTML]{F4CCCC}Universally missing & ['None'], &  \\
Adjusting to the environment & providing light & led light & \cellcolor[HTML]{F4CCCC}Universally missing & ['None'], & led bulb \\
Adjusting to the environment & ventilating & ventilation hole & \cellcolor[HTML]{F4CCCC}Universally missing & ['None'], & vent \\
Adjusting to the environment & providing heat & furnace & \cellcolor[HTML]{F4CCCC}Universally missing & ['None'], & fire \\
Adjusting to the environment & ventilating & vent pipe & \cellcolor[HTML]{F4CCCC}Universally missing & ['None'], &  \\
Adjusting to the environment & covering floors & floor mat & \cellcolor[HTML]{F4CCCC}Universally missing & ['None'], &  \\
Adjusting to the environment & draining waste liquid & bath drain & \cellcolor[HTML]{F4CCCC}Universally missing & ['None'], & drain/risks loss of granularity \\
Beautifying & enhancing body appearance & fragrance & \cellcolor[HTML]{F4CCCC}Universally missing & ['None'], & perfume \\
Beautifying & adorning the body & bindi & \cellcolor[HTML]{F4CCCC}Universally missing & ['None'], &  \\
Cooking & providing heat for cooking & clay stove & \cellcolor[HTML]{F4CCCC}Universally missing & ['None'], & stove/risks loss of granularity \\
Cooking & providing heat for cooking & charcoal stove & \cellcolor[HTML]{F4CCCC}Universally missing & ['None'], & stove/risks loss of granularity \\
Cooking & providing heat for cooking & brick oven & \cellcolor[HTML]{F4CCCC}Universally missing & ['None'], &  \\
Cooking & holding food during cooking & clay pot & \cellcolor[HTML]{F4CCCC}Universally missing & ['None'], &  \\
Cooking & gathering and storing drinking water & jerrycan & \cellcolor[HTML]{F4CCCC}Universally missing & ['None'], &  \\
Cooking & providing heat for cooking & stone stove & \cellcolor[HTML]{F4CCCC}Universally missing & ['None'], & stove/risks loss of granularity \\
Cooking & storing perishable food & earthern pot & \cellcolor[HTML]{F4CCCC}Universally missing & ['None'], &  \\
Healing & providing medication & ointment & \cellcolor[HTML]{F4CCCC}Universally missing & ['None'], &  \\
Healing & providing medication & antiseptic & \cellcolor[HTML]{F4CCCC}Universally missing & ['None'], &  \\
Healing & providing medication & balm & \cellcolor[HTML]{F4CCCC}Universally missing & ['None'], &  \\
Healing & providing medication & box of medication & \cellcolor[HTML]{F4CCCC}Universally missing & ['None'] & medication \\
Healing & providing medication & medicinal liquid & \cellcolor[HTML]{F4CCCC}Universally missing & ['None'], &  \\
Maintaining a diurnal cycle & sleeping & charpai & \cellcolor[HTML]{F4CCCC}Universally missing & ['None'], &  \\
Practicing hygiene & cleaning teeth & tooth soap & \cellcolor[HTML]{F4CCCC}Universally missing & ['None'], &  \\
Practicing hygiene & washing clothes & laundry pod & \cellcolor[HTML]{F4CCCC}Universally missing & ['None'], &  \\
Practicing hygiene & absorbing menstrual flow & panty liner & \cellcolor[HTML]{F4CCCC}Universally missing & ['None'], &  \\
Practicing hygiene & shaving & shaving foam & \cellcolor[HTML]{F4CCCC}Universally missing & ['None'], &  \\
Practicing hygiene & bathing & bamboo enclosure & \cellcolor[HTML]{F4CCCC}Universally missing & ['None'], &  \\
Practicing hygiene & cleaning teeth & tooth powder & \cellcolor[HTML]{F4CCCC}Universally missing & ['None'], &  \\
Practicing hygiene & washing clothes & washboard & \cellcolor[HTML]{F4CCCC}Universally missing & ['None'], &  \\
Practicing hygiene & collecting garbage & landfill & \cellcolor[HTML]{F4CCCC}Universally missing & ['None'], &  \\
Practicing hygiene & cleaning teeth & clay/brick powder & \cellcolor[HTML]{F4CCCC}Universally missing & ['None'], &  \\
Practicing hygiene & drying clothes & tree trunk & \cellcolor[HTML]{F4CCCC}Universally missing & ['None'], &  \\
Adjusting to the environment & seating & bean bag chair & \cellcolor[HTML]{FFF2CC}Single-dataset & ['Open Images'], & bean bag \\
Adjusting to the environment & draining waste liquid & sinkhole & \cellcolor[HTML]{FFF2CC}Single-dataset & ['Open Images'], &  \\
Adjusting to the environment & draining waste liquid & cistern & \cellcolor[HTML]{FFF2CC}Single-dataset & ['Lvis'], &  \\
Adjusting to the environment & covering floors & planks & \cellcolor[HTML]{FFF2CC}Single-dataset & ['Open Images'], &  \\
Adjusting to the environment & controlling access & keycard & \cellcolor[HTML]{FFF2CC}Single-dataset & ['Lvis'], &  \\
Adjusting to the environment & providing shelter & shed & \cellcolor[HTML]{FFF2CC}Single-dataset & ['Open Images'], &  \\
Adjusting to the environment & providing shelter & cabin & \cellcolor[HTML]{FFF2CC}Single-dataset & ['Open Images'], &  \\
Adjusting to the environment & providing heat & gas stove & \cellcolor[HTML]{FFF2CC}Single-dataset & ['Open Images'], & stove/ risks loss of granularity \\
Adjusting to the environment & seating & mat & \cellcolor[HTML]{FFF2CC}Single-dataset & ['Open Images'], &  \\
Adjusting to the environment & cooling living spaces & hand fan & \cellcolor[HTML]{FFF2CC}Single-dataset & ['Open Images'], &  \\
Adjusting to the environment & seating & massage chair & \cellcolor[HTML]{FFF2CC}Single-dataset & ['Open Images'], &  \\
Adjusting to the environment & providing shelter & tent & \cellcolor[HTML]{FFF2CC}Single-dataset & ['Open Images'], &  \\
Beautifying & enhancing body appearance & mascara & \cellcolor[HTML]{FFF2CC}Single-dataset & ['Open Images'], &  \\
Beautifying & decorating walls & collage & \cellcolor[HTML]{FFF2CC}Single-dataset & ['Open Images'], &  \\
Beautifying & decorating walls & potted plant & \cellcolor[HTML]{FFF2CC}Single-dataset & ['COCO'], & 
\end{tabular}%
}
\end{table*}

\begin{table*}[ht]
\centering
\resizebox{\textwidth}{!}{%
\begin{tabular}{c|c|c|c|c|c}
\hline
\textbf{\begin{tabular}[c]{@{}c@{}}Human Universal \\ Affordance\end{tabular}} & \textbf{Funtional Name} & \textbf{Objects} & \textbf{Coverage} & \textbf{Datasets Covered} & \textbf{Notes/Possible rephrase} \\ \hline
Beautifying & adorning the body & rosary & \cellcolor[HTML]{FFF2CC}Single-dataset & ['Open Images'], &  \\
Beautifying & decorating walls & mural & \cellcolor[HTML]{FFF2CC}Single-dataset & ['Open Images'], &  \\
Beautifying & decorating walls & statue & \cellcolor[HTML]{FFF2CC}Single-dataset & ['Open Images'], &  \\
Beautifying & enhancing body appearance & eyeliner & \cellcolor[HTML]{FFF2CC}Single-dataset & ['Open Images'], &  \\
Cooking & storing perishable food & rack & \cellcolor[HTML]{FFF2CC}Single-dataset & ['Open Images'], &  \\
Healing & providing medication & inhaler & \cellcolor[HTML]{FFF2CC}Single-dataset & ['Lvis'], &  \\
Healing & providing rehabilitation support & cane & \cellcolor[HTML]{FFF2CC}Single-dataset & ['Open Images'], &  \\
Healing & providing medication & leaves & \cellcolor[HTML]{FFF2CC}Single-dataset & ['Open Images'], &  \\
Practicing hygiene & washing clothes & rock & \cellcolor[HTML]{FFF2CC}Single-dataset & ['Open Images'], &  \\
Practicing hygiene & cleaning household & rake & \cellcolor[HTML]{FFF2CC}Single-dataset & ['Open Images'], &  \\
Practicing hygiene & washing clothes & brush & \cellcolor[HTML]{FFF2CC}Single-dataset & ['Open Images'], &  \\
Practicing hygiene & bathing & river & \cellcolor[HTML]{FFF2CC}Single-dataset & ['Open Images'], &  \\
Practicing hygiene & bathing & faucet & \cellcolor[HTML]{FFF2CC}Single-dataset & ['Lvis'], &  \\
Practicing hygiene & hand washing & hand sanitizer & \cellcolor[HTML]{FFF2CC}Single-dataset & ['Open Images'], &  \\
Practicing hygiene & removing dirt from dishes & ashes & \cellcolor[HTML]{FFF2CC}Single-dataset & ['Open Images'], &  \\
Practicing hygiene & absorbing menstrual flow & tampon & \cellcolor[HTML]{FFF2CC}Single-dataset & ['Open Images'], &  \\
Practicing hygiene & bathing & water well & \cellcolor[HTML]{FFF2CC}Single-dataset & ['Open Images'], & well \\
Practicing hygiene & bathing & water tank & \cellcolor[HTML]{FFF2CC}Single-dataset & ['Open Images'], & tank \\
Traveling & transporting & scooter & \cellcolor[HTML]{FFF2CC}Single-dataset & ['Open Images'], &  \\
Traveling & transporting & van & \cellcolor[HTML]{FFF2CC}Single-dataset & ['Open Images'], &  \\
Adjusting to the environment & ventilating & screen & \cellcolor[HTML]{D9EAD3}Multi-dataset & ['ImageNet', 'Open Images'], &  \\
Adjusting to the environment & seating & rocking chair & \cellcolor[HTML]{D9EAD3}Multi-dataset & ['ImageNet', 'Open Images', 'Lvis'], &  \\
Adjusting to the environment & controlling access & latch & \cellcolor[HTML]{D9EAD3}Multi-dataset & ['Open Images', 'Lvis'], &  \\
Adjusting to the environment & ventilating & air conditioner & \cellcolor[HTML]{D9EAD3}Multi-dataset & ['Lvis', 'Open Images'], &  \\
Adjusting to the environment & draining waste liquid & toilet & \cellcolor[HTML]{D9EAD3}Multi-dataset & ['Open Images', 'Lvis', 'COCO', 'Dollar Street'], &  \\
Adjusting to the environment & providing heat & boiler & \cellcolor[HTML]{D9EAD3}Multi-dataset & ['Open Images', 'YFCC Entity'], &  \\
Adjusting to the environment & providing shelter & yurt & \cellcolor[HTML]{D9EAD3}Multi-dataset & ['ImageNet', 'Open Images'], &  \\
Beautifying & decorating walls & figurine & \cellcolor[HTML]{D9EAD3}Multi-dataset & ['Open Images', 'Lvis'], &  \\
Beautifying & decorating walls & curtain & \cellcolor[HTML]{D9EAD3}Multi-dataset & ['Open Images', 'Lvis'], &  \\
Beautifying & decorating walls & portrait & \cellcolor[HTML]{D9EAD3}Multi-dataset & ['Open Images', 'Lvis', 'Dollar Street'], &  \\
Cooking & preparing food ingredients & mallet & \cellcolor[HTML]{D9EAD3}Multi-dataset & ['Open Images', 'Lvis'], &  \\
Cooking & preparing food ingredients & paddle & \cellcolor[HTML]{D9EAD3}Multi-dataset & ['ImageNet', 'Open Images', 'Lvis'], &  \\
Cooking & storing perishable food & basket & \cellcolor[HTML]{D9EAD3}Multi-dataset & ['Open Images', 'Lvis'], &  \\
Cooking & providing space for eating & stool & \cellcolor[HTML]{D9EAD3}Multi-dataset & ['Open Images', 'Lvis'], &  \\
Cooking & gathering and storing drinking water & calabash & \cellcolor[HTML]{D9EAD3}Multi-dataset & ['Open Images'], &  \\
Cooking & preparing food ingredients & blender & \cellcolor[HTML]{D9EAD3}Multi-dataset & ['Open Images', 'Lvis'], &  \\
Cooking & preparing food ingredients & chopstick & \cellcolor[HTML]{D9EAD3}Multi-dataset & ['Open Images', 'Lvis'], &  \\
Cooking & providing heat for cooking & hot plate & \cellcolor[HTML]{D9EAD3}Multi-dataset & ['Open Images', 'Lvis'], &  \\
Healing & providing medication & lotion & \cellcolor[HTML]{D9EAD3}Multi-dataset & ['ImageNet', 'Open Images', 'Lvis'], &  \\
Maintaining a diurnal cycle & sleeping & sleeping bag & \cellcolor[HTML]{D9EAD3}Multi-dataset & ['ImageNet', 'Open Images', 'Lvis'], &  \\
Maintaining a diurnal cycle & sleeping & bedspread/bedding & \cellcolor[HTML]{D9EAD3}Multi-dataset & ['Lvis', 'Open Images'], &  \\
Maintaining a diurnal cycle & waking up and keeping time & alarm clock & \cellcolor[HTML]{D9EAD3}Multi-dataset & ['Open Images', 'Lvis', 'Dollar Street'], &  \\
Practicing hygiene & washing clothes & barrel & \cellcolor[HTML]{D9EAD3}Multi-dataset & ['ImageNet', 'Open Images', 'Lvis'], &  \\
Practicing hygiene & removing body dirt (substances) & alcohol & \cellcolor[HTML]{D9EAD3}Multi-dataset & ['Open Images', 'Lvis'], &  \\
Practicing hygiene & cleaning teeth & salt & \cellcolor[HTML]{D9EAD3}Multi-dataset & ['Open Images', 'Dollar Street'], &  \\
Practicing hygiene & cleaning body after toilet use & newspaper & \cellcolor[HTML]{D9EAD3}Multi-dataset & ['Open Images', 'Dollar Street'], &  \\
Practicing hygiene & cleaning household & sponge & \cellcolor[HTML]{D9EAD3}Multi-dataset & ['Open Images', 'Lvis'], &  \\
Practicing hygiene & hand washing & kettle & \cellcolor[HTML]{D9EAD3}Multi-dataset & ['Open Images', 'Lvis'], &  \\
Practicing hygiene & hand washing & hose & \cellcolor[HTML]{D9EAD3}Multi-dataset & ['Open Images', 'Lvis'], &  \\
Practicing hygiene & hand washing & fountain & \cellcolor[HTML]{D9EAD3}Multi-dataset & ['ImageNet', 'Open Images'], &  \\
Traveling & transporting & cart & \cellcolor[HTML]{D9EAD3}Multi-dataset & ['Open Images', 'Lvis'], & 
\end{tabular}%
}
\caption{List of 100 longtail objects from Culture Affordance Atlas with details regarding coverage in popular VL datasets}
\label{tab:longtail_details2}
\end{table*}

\subsection{Function-Topic Image Figures}

\begin{figure}[ht]
    \centering
    \includegraphics[width=0.70\columnwidth]{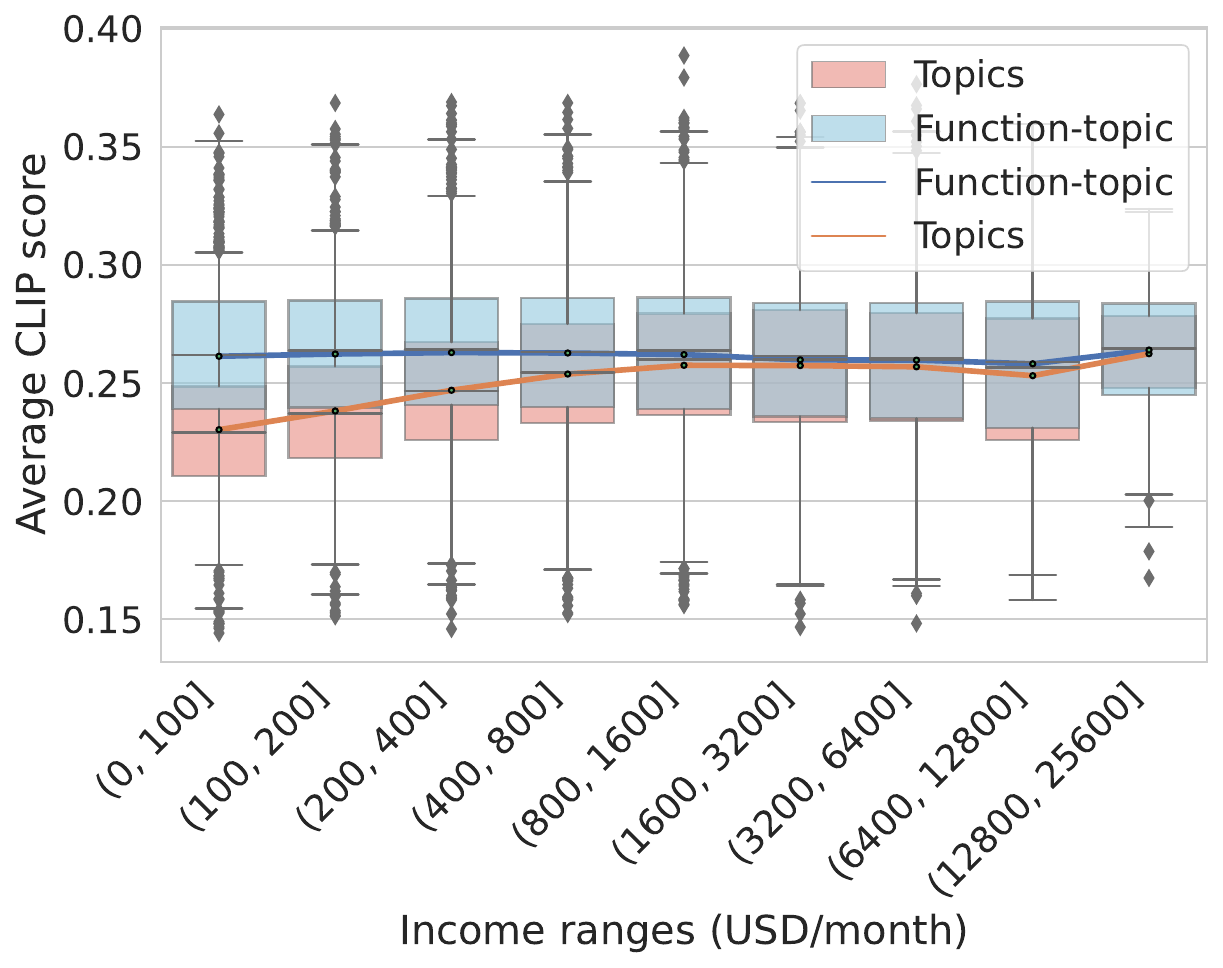}
    \caption{Comparison of CLIP alignment scores between Topic-Image (red) and Function\_Topic-Image (blue) pairs across Dollar Street images from various income levels. Trend lines represent mean scores per income bin. The slope comparison illustrates how function-centric labeling helps reduce the digital divide. \textit{Best viewed in color.}}
    
    \label{fig:TopicvsFunctionTopicBox}
\end{figure}

\begin{figure}[ht]
    \centering
    \includegraphics[width=0.9\columnwidth]{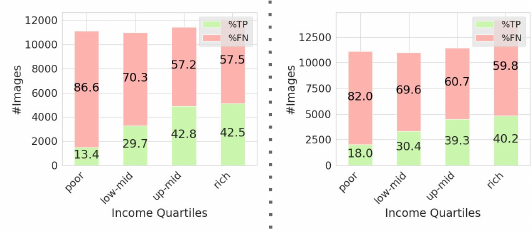}
    \caption{CLIP Recall across all images using Topic-Image (left) and Function\_Topic-Image (right) alignment scores. We report the percentage of true positives  and false negatives for each income quartile. Function\_Topic-Image Recall exhibits less variation across income levels compared to Topic-Image Recall, suggesting improved consistency and reduced income-related bias. \textit{Best viewed in color.}}
    
    \label{fig:TopicvsFunctionTopicBar}
\end{figure}

\begin{figure}[ht]
    \centering
    \includegraphics[width=0.70\columnwidth]{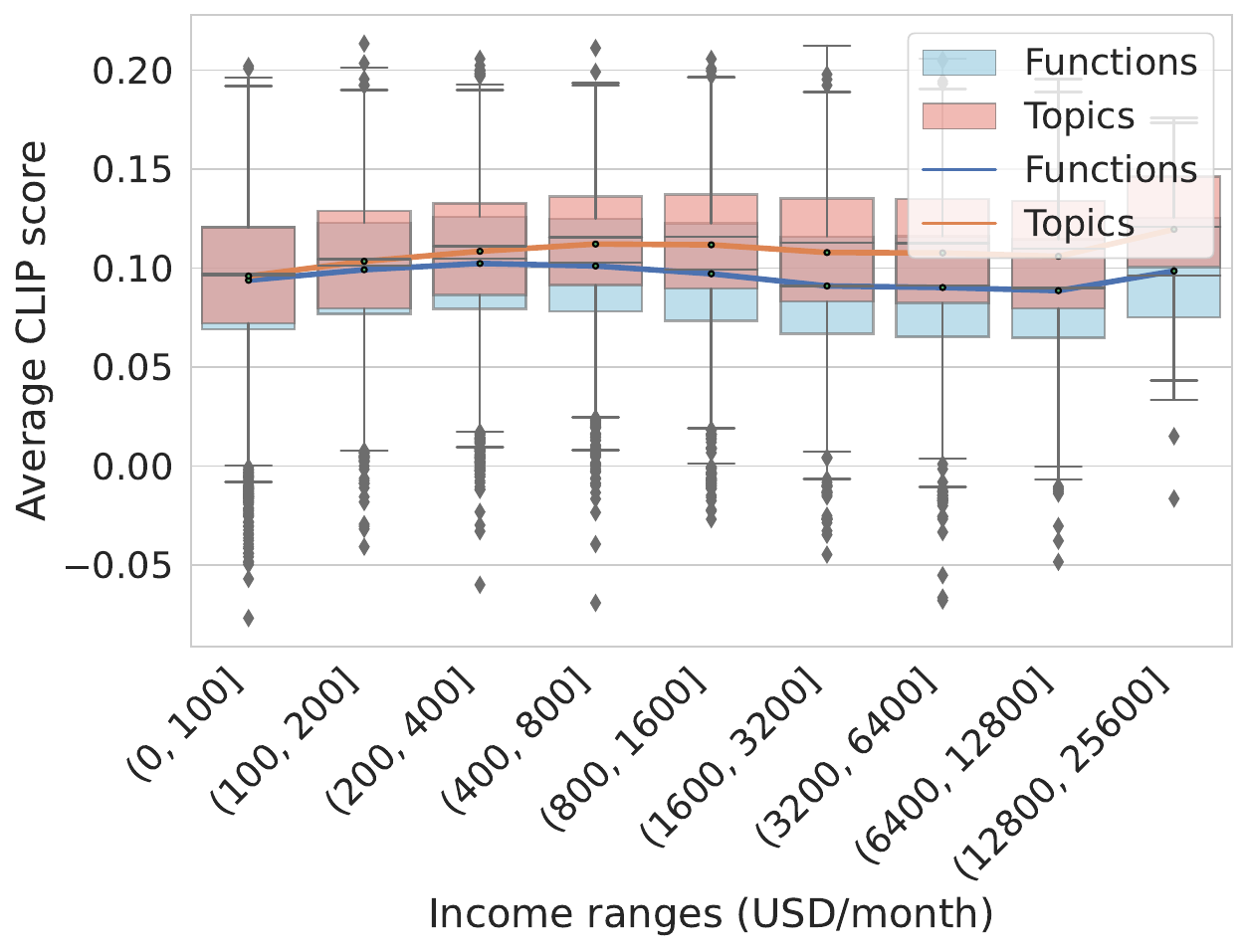}
    \caption{Comparison of SigLIP2 alignment scores between Topic-Image (red) and Function-Image (blue) pairs across Dollar Street images from varying income levels. Trend lines indicate mean scores per income bin. The slope of each line reflects the extent of the digital divide. \textit{Best viewed in color.}}
    
    \label{fig:siglipTopicvsFunctionBox}
\end{figure}

\begin{figure}[ht]
    \centering
    \includegraphics[width=0.90\columnwidth]{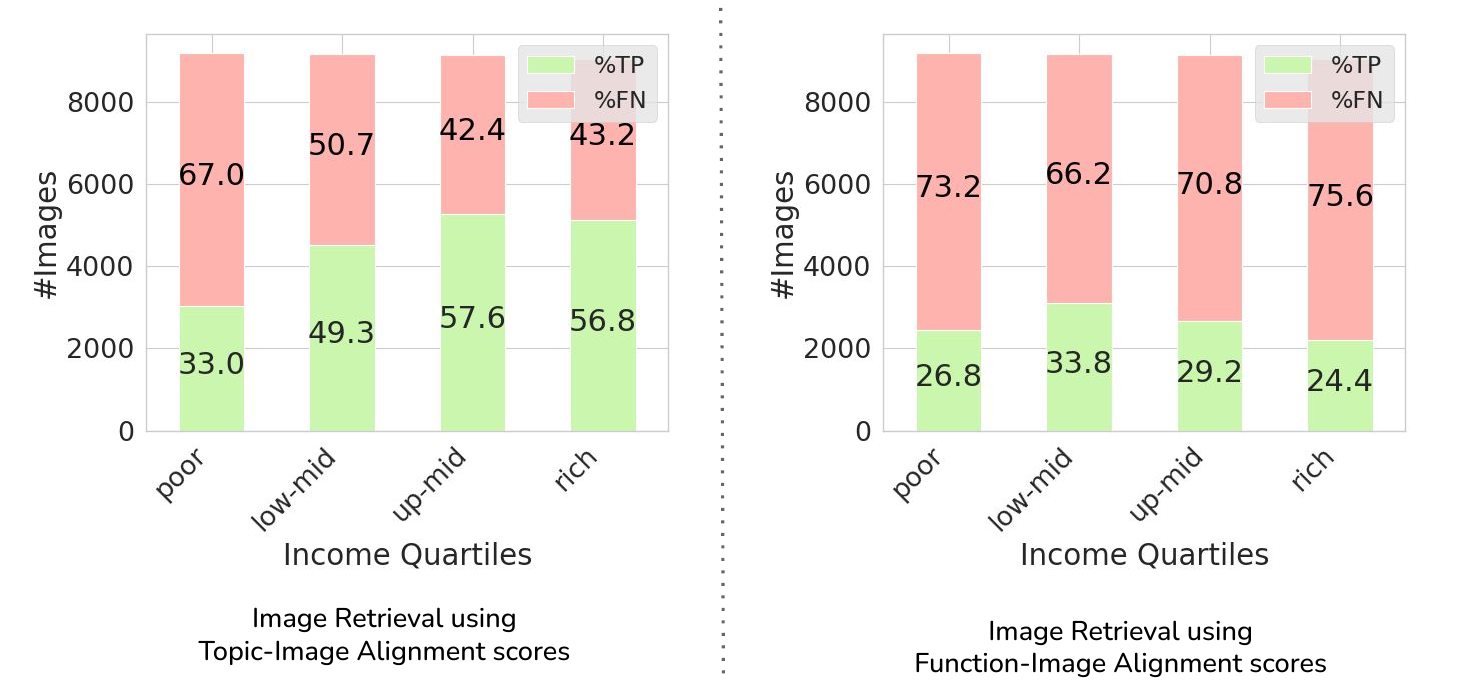}
    \caption{SigLIP2 Recall across all images using Topic-Image (left) and Function-Image (right) alignment scores. We report the percentage of true positives (``recognized'' images) and false negatives (``forgotten'' images) for each income quartile. Function-Image Recall shows less variation across income levels compared to Topic-Image Recall, indicating greater robustness to income-based distribution shifts. \textit{Best viewed in color.}}
    \label{fig:siglipTopicvsFunctionBar}
\end{figure}

\begin{table*}[ht]
\centering
\resizebox{0.70\textwidth}{!}{%
\begin{tabular}{l|l|l|l|l}
\textbf{Country} & \textbf{\% Agreement} & \textbf{$ \kappa/\alpha $} & \textbf{Gwet's AC1} & \textbf{PABAK} \\ \hline
China         & 77 & 0.28 (Fleiss)       & 0.8  & 0.69 \\
India         & 83 & 0.51 (Fleiss)       & 0.89 & 0.82 \\
Ethiopia      & 83 & 0.31 (Fleiss)       & 0.87 & 0.78 \\
Nigeria       & 70 & 0.04 (Fleiss)       & 0.71 & 0.56 \\
United States & 77 & 0.38 (Fleiss)       & 0.83 & 0.73 \\
Romania       & 90 & -0.03 (Fleiss)      & 0.93 & 0.87 \\
Russia        & 80 & 0.19 (Cohen)        & 0.79 & 0.67 \\ \hline
\textbf{Overall}       & \textbf{80} & \textbf{0.27 (Krippendorff)} & \textbf{0.83} & \textbf{0.73}
\end{tabular}%
}
\caption{Inter-annotator agreement metrics. While raw agreement was high across countries (70–90\%, avg. 80\%), traditional metrics like Fleiss’ Kappa show variability due to class imbalance. In contrast, Gwet’s AC1 (0.83) and PABAK (0.73) indicate substantial overall agreement, suggesting strong annotator alignment despite cultural nuances.
}
\label{tab:interannotator}
\end{table*}

\end{document}